# Eliminating the spin-down critical angle in polarizing neutron optics for expanding the polarization bandwidth

A. Zubayer,[1] N. Ghafoor,[1] A. Glavic,[2] J. Stahn,[2] M. Lorentzon[1], A. Le Febvrier,[1] P. Eklund,[1] J. Birch,[1] F. Eriksson[1]

1. Department of Physics, Chemistry, and Biology, IFM, Linköping University, SE-581 83 Linköping, Sweden

2. Laboratory for Neutron Scattering and Imaging, Paul Scherrer Institut, 5232, Villigen PSI, Switzerland

**Polarized neutron scattering is a very important analysis technique for studies of magnetism, spintronics, and high-sensitivity measurements, among others, offering invaluable information. Yet, the efficiency of such experiments rely on the performance of the polarizing neutron optics to provide high reflectivity and polarization. Presently, state-of-the-art polarizers like Fe/Si supermirrors are not able to polarize neutrons at low scattering angles in a monochromatic beam or able to polarize neutrons with a variety of wavelengths as for a non-monochromatic beam. To overcome this limitation, it is suggested to use Co/Ti multilayers on Ti substrates owing to their favorable scattering length density characteristics. It is shown that this approach enables a wavelength bandwidth several times larger than achievable with state-of-the-art materials on Si or glass substrates. Consequently, enabling the possibility of polarizing neutrons across an extended wavelength range, including those with very high wavelengths.**

Neutron scattering, which is an inherently contrained technique with its limited signal and time-intensive nature, can be significantly improved through advancements in neutron optics, particularly the neutron mirrors.[1] Polarized neutron experiments, crucial for studies of magnetism, spintronics, and high-sensitivity measurements, among others, rely on selecting either spin-up or spin-down states to pass through. However, current state-of-the-art materials for polarizing neutron optics exhibit limitations in their polarization bandwidth. Nuclear-reactor sources produce monochromatic beams, while spallation sources emit non-monochromatic beams with a wide range of neutron wavelengths. For a monochromatic beam, different critical angles exists for spin-up and spin-down neutrons, with polarization achievable only above the spin-down critical angle due to the total external reflection below this angle. One limit of current state-of-the-art material systems for polarizing neutron optics, like Fe/Si, is the polarization capability for large incidence angles.[2–6] However, there is also a large non-polarized range at small incidence angles, below the spin-down critical angle, where the intensity is the highest.[7]

For a non-monochromatic beam, the same limitation applies due to the inability to polarize neutrons with large wavelengths.

Here we propose a novel approach to achieve polarization within this low-angle region for a monochromatic beam, thus also extending the non-polarized region for high-wavelength neutrons in a non-monochromatic beam. Our approach utilizes the low scattering length density (SLD) of Ti in a combination of Co/Ti multilayers on a Ti substrate, replacing Si or glass. This configuration effectively eliminates the critical angle for spin down, enabling a polarization from 0° incidence angle

Polarization in neutron optics relies on both material selection and geometry. By depositing a substrate with alternating layers of magnetic and non-magnetic materials,[8] the neutron interactions are affected based on the neutron spin, correlated to the SLD. This SLD depends on the scattering length and mass density of the atoms,[9] playing a crucial role for reflection at interfaces between different media. Neutrons have distinct scattering behaviour in magnetic layers depending on their spin orientation. This results in scattering length densities where the magnetic scattering length density is either added or subtracted from the nuclear scattering length density, for spin up and down, respectively. To achieve optimal polarization, the SLD for spin-down neutrons in magnetic layers must have the same value asthe SLD of non-magnetic layers, while for spin-up neutrons the difference should be as large as possible. Material systems such as Fe/Si and Co/Ti are preferred for their matching SLDs and have been extensively studied for polarizing optics.[10–16]

Enabling polarization also in the low-angle region enhances the efficiency of utilizing available neutrons in a monochromatic beam. On the contrary, in a non-monochromatic beam, the polarized wavelength bandwidth, $\lambda_{max}$, follows the following equation:

$$\lambda_{max} = \lambda_{min} \cdot \left(\frac{Q_m}{Q_{min-polarized}}\right) \qquad (1)$$

Here $\lambda_{min}$ is the smallest wavelength within the beam, $Q_{max}$ is the highest Q-value of reflection and polarization, and $Q_{min-polarized}$ denotes the lowest scattering vector for polarization. The scattering vector, Q, which depends on both the wavelength and the scattering angle,is:

$$Q = \frac{4\pi}{\lambda} \sin\left(\frac{2\theta}{2}\right). \qquad (2)$$

This equation also shows that a low scattering angle corresponds to a low scattering vector. According to Equation (1), it implies that achieving polarization at low angles, or low scattering vectors,for a monochromatic beam significantly increase the polarized bandwidth for a non-monochromatic beam.

Previous attempts to achieve low-angle polarization in neutron optics encountered significant challenges. One method involved directing neutrons through the substrate before reflecting off the multilayer from below, but this significantly reduced the achievable m-value [ref]. Another approach used a Gd absorbing layer beneath the multilayer to address the spin-down critical angle.[17] However, this method prevents transmission mode usage, and with a requirement of the Gd layer to be over 1000 Å thick the mirror smoothness was substantially affeced. In addition, the neutron interaction with Gd generates hard gamma radiation requiring extensive shielding.

Our methodology focuses on material selection to reduce or eliminate the spin-down critical angle, which is crucial since the critical angle depends on the average scattering length density (SLD) encountered by neutrons, as seen in Equation (4) in Appendix. The SLD of the substrate plays an decisive role for setting the lower limit for the sample's critical angle. Therefore, choosing multilayers and substrates with a low average SLD is essential for minimizing this angle. Ti and Mn have the lowest SLD values, with Ti being preferred due to its extensive use in neutron optics[18] and less radiation concerns compared to Mn. Replacing Si substrates with Ti can effectively eliminate the spin-down critical angle. Furthermore, adjusting the layer thickness ratio between magnetic and non-magnetic layers allows for reaching higher critical angles for spin-up neutrons while at the same time reducing the spin-down critical angle. Co/Ti multilayers on a Ti substrate offer a smaller total SLD than Fe/Si, providing an optimal solution for polarization. The preference of Co/Ti multilayers on Ti over Fe/Si on Ti, is based on the lower SLD of the Co/Ti multilayer. This study evaluates different layer thickness ratios of Co/Ti multilayers on Ti substrates in comparison to Fe/Si on Si and Co/Ti on Si to determine the most effective polarization configuration.

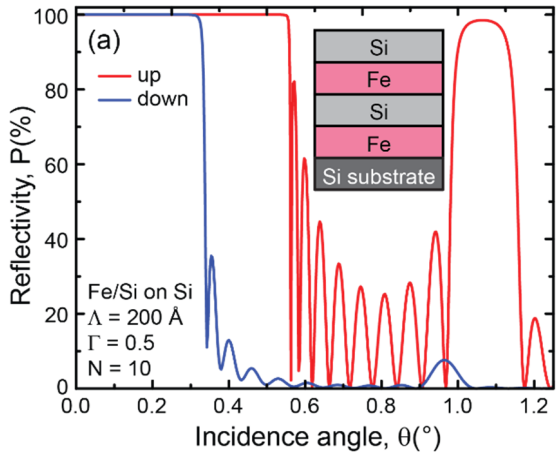 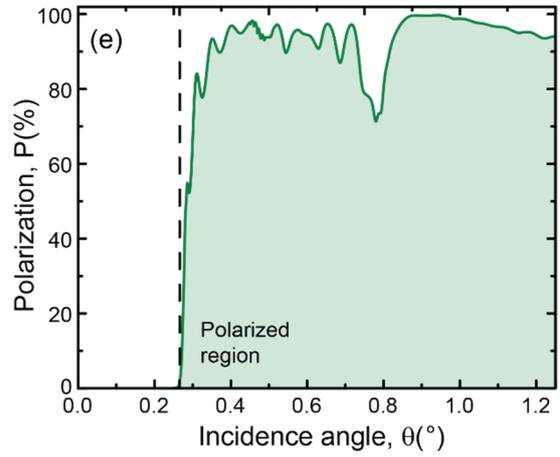
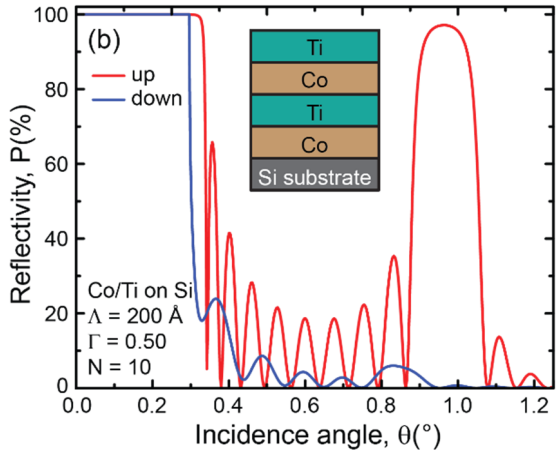 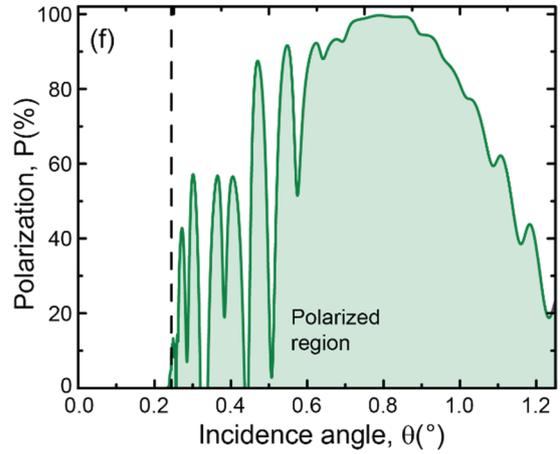
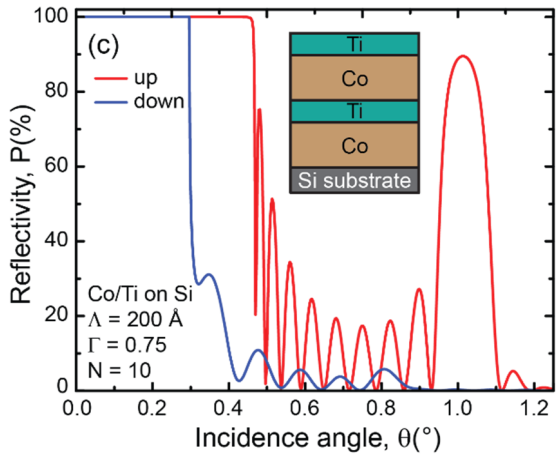 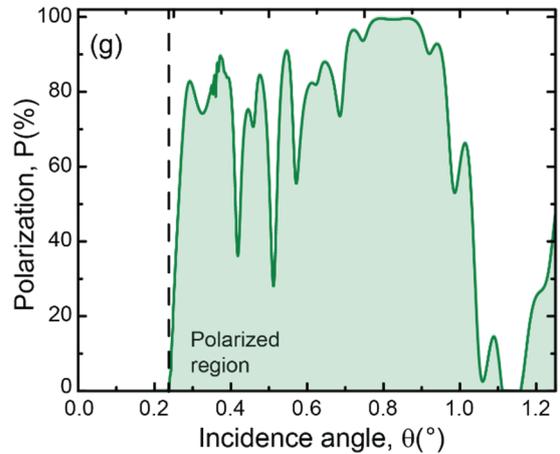
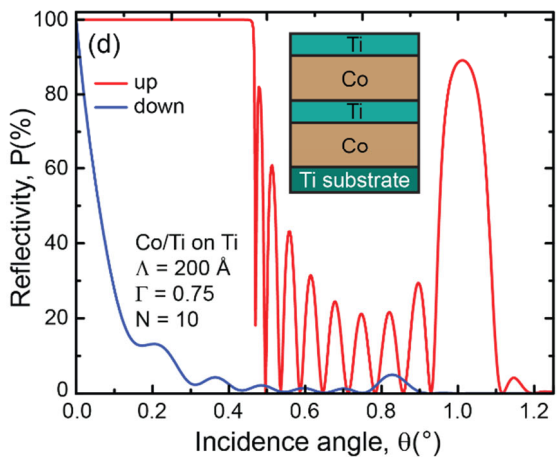 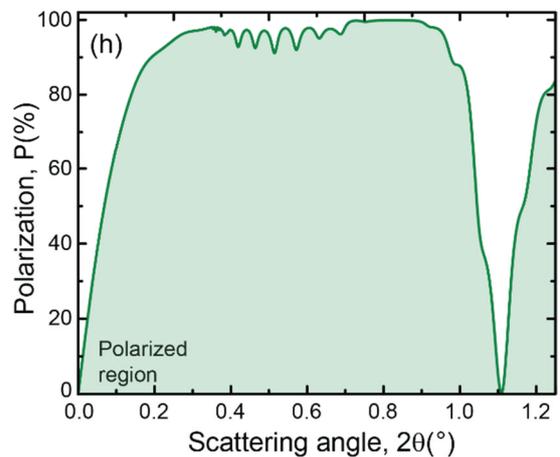

**Figure 1. Polarized neutron reflectivity simulations.** Schematic illustrations of the multilayer structures and the corresponding polarized neutron reflectivity simulations, showing the non-polarized (yellow) and polarized (pink) regions below the spin-down and spin-up critical angles, respectively. The number of periods is N = 10 and a period thickness of $\Lambda$ = 200 Å. The wavelength is 5 Å since this is a common wavelength for neutron reflectometers. (a) Fe/Si ($\Gamma$ = 0.5) on Si, (b) Co/Ti ($\Gamma$ = 0.5) on Si, (c) Co/Ti ($\Gamma$ = 0.75) on Si and (d) Co/Ti ($\Gamma$ = 0.75) on Ti. (e-f) shows the polarization curves for (a-d).

Figure 1 shows polarized neutron reflectivity simulations of 4 different multilayer samples. The red curves represent the spin-up reflectivity while the blue curves the spin-down reflectivity. The region below the spin-down angle is coloured yellow. In this region, we have the total external reflectivity for both spin-up and spin-down neutrons. Between the spin-down and spin-up critical angles we have the total external reflectivity for the spin-up neutrons only, hence that region is polarized and is represented with a pink background. The region above the spin-up critical angle is also polarizable depending on the multilayer period thicknesses and is thus also represented with a pink background. Our investigation is to widen the polarized region below the spin-up critical angle, by diminishing the spin-down critical angle. In the inset of each subfigure in Figure 1, we see the corresponding multilayer for each simulated PNR. Note however that the period number for the simulation is 10 and not 3 as it shows in the insets. Figure 1(a) shows the PNR simulations of Fe/Si on Si. When the multilayer system is Co/Ti on Si, as seen in Figure 1(b), the spin-down and spin-up critical angles decreased. Further, when changing the layer thickness ratio of Co/Ti on Si to 0.75, as seen in Figure 1(c) the spin-down critical angle remained the same, but the spin-up critical angle increased. The final change is seen in Figure 1(d), where the Si substrate was exchanged with Ti, and the spin-down critical angle was eliminated.

The spin-up critical angle increase for a higher thickness ratio ($\Gamma$) of magnetic layer can be seen by comparing the simulations in subfigure (b) and (c), but also seen in Figure S3. The complete elimination of the spin-down critical angle by exchanging Si substrate with Ti substrate is obvious from the simulations in Figure (d). Samples corresponding to (a), (c) and (d) were grown and measured to experimentally compare with the simulations.

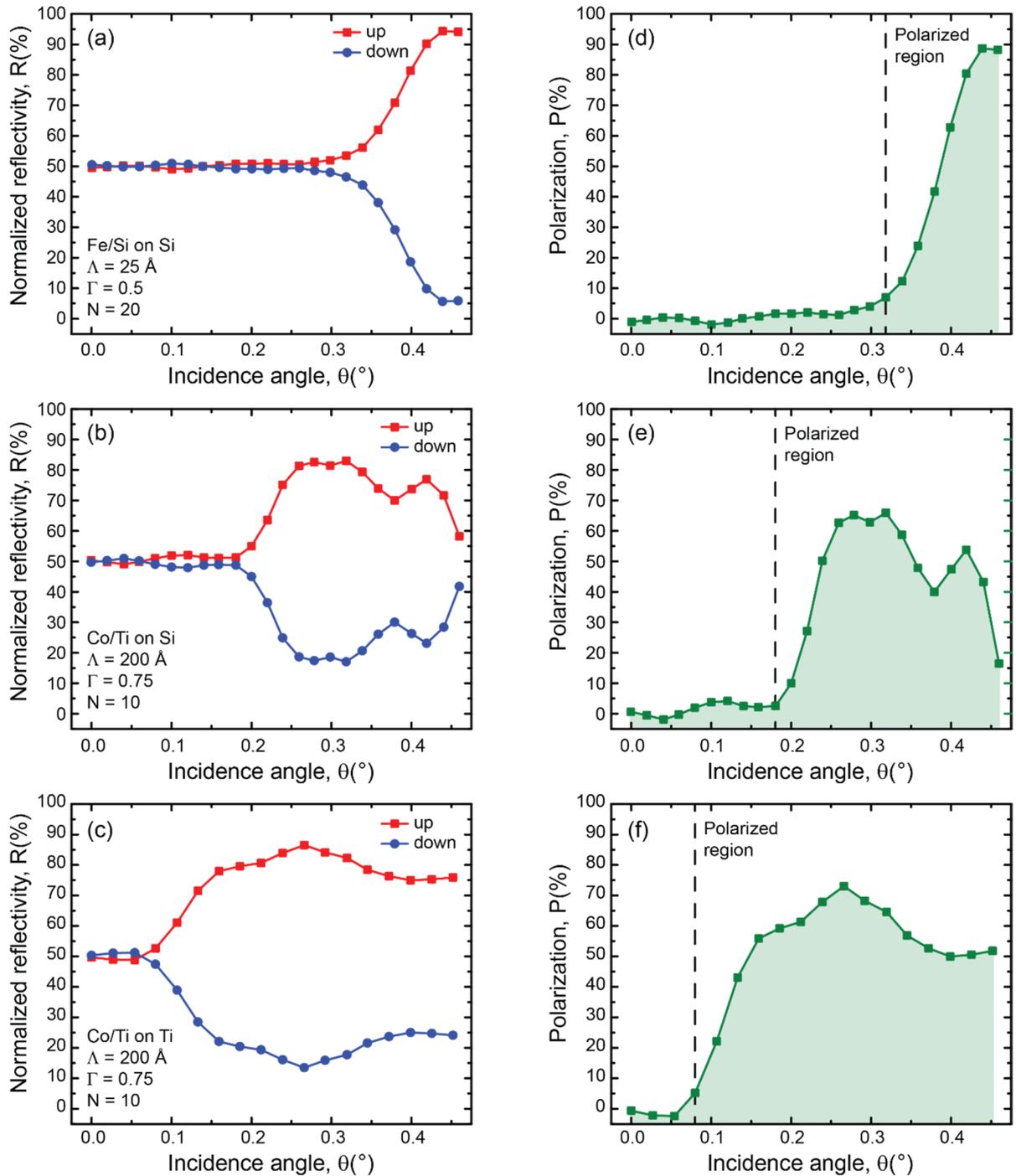

**Figure 2. Neutron polarization measurements.** (a-c) normalized intensity of reflected neutron of the two spins for Fe/Si on Si (Λ = 25 Å, N=20, Γ = 0.5), Co/Ti on Si (Λ = 200 Å, N=10, Γ = 0.75) and Co/Ti on Ti (Λ = 200 Å, N=10, Γ = 0.75) multilayers, respectively. (d-f) polarization of (a-c) respective.

Figure 2 shows the polarization of Fe/Si on Si (Γ = 0.5), Co/Ti on Si (Γ = 0.75) and Co/Ti on Ti (Γ = 0.75) over angles between 0 and 0.6 degrees in 2θ. The polarization curves represent the ratio of spin-up (red) and spin-down (blue) neutrons that are being reflected at various angles. The polarized region has a pink background while the non-polarized region is yellow. It is important to note before interpreting the data from Figure 2 that experimentally it is impossible to fully mask the direct beam to investigate the polarization at 0. However, it is possible to reduce the direct beam region by decreasing the size of the beam that illuminates the sample. What we lose in intensity we will gain in resolution and achieve a smaller direct beam region. The direct beam is always non-polarized, thus for very small angles (below approximately 0.18 in 2θ) parts of the beam directly hit the detector with a very high intensity with non-polarized neutrons. This leads to a direct beam response below 0.18 in 2θ for all measurements.

The Fe/Si on Si sample, seen in Figure 2(a), has the largest non-polarized region due to the spin up and down critical angles being at high scattering angles. The Co/Ti on Si sample, seen in Figure 2(b), has a smaller non-polarized region due to the lower spin-up critical angle. Hence, as per our hypothesis, the lower the SLD that the neutrons' path undergoes, the lower the critical angles/vectors. Finally, the Co/Ti on Ti sample, as seen in Figure 2(c), completely eliminated the non-polarized region, meaning that as soon as the direct beam ends, there is polarization.

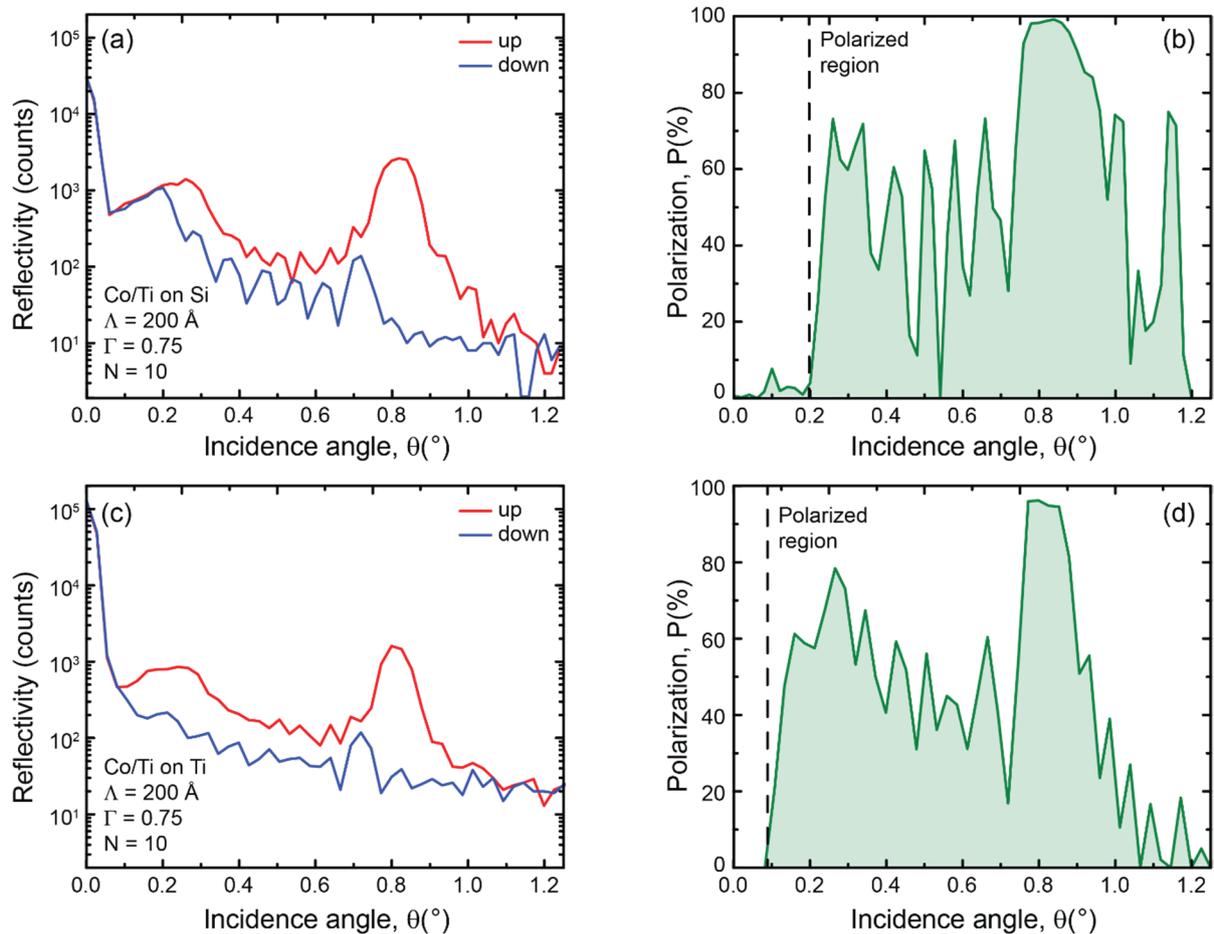

**Figure 3. Polarized neutron reflectivity measurements.** PNR of Co/Ti on Si (Λ = 200 Å, N=10, Γ = 0.75) (a) and Co/Ti on Ti (Λ = 200 Å, N=10, Γ = 0.75) (b).

Although the ideal scenarios have been described the real-life scenario must include the direct beam at the very lowest angles where part of the beam travels above the multilayer itself. Further, the beam is a gaussian and thus the shape of the critical edge will follow a rounder shape. Therefore, the importance of this study is not the reflectivity intensity comparisons between the material systems but rather the polarization and the concept of spin-down critical angle elimination.

Comparing the polarized neutron reflectivity (PNR) of Co/Ti on Si (Figure 3a) with Co/Ti on Ti (Figure 3b) we can in a wider scattering range visualize and compare the critical edges of both samples. It was also evident with PNR, just as with XRR seen in Figure S2 in the Appendix, that the rough Ti substrate, compared to Si, significantly decreases the intensity. This could be compensated for, with a Ti buffer layer investigated in the Appendix.

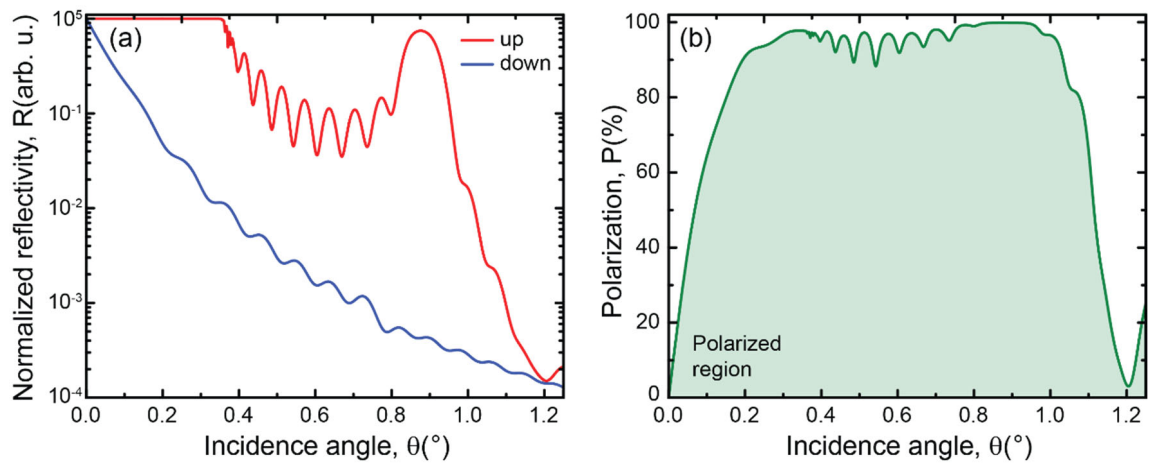

**Figure 4. Simulation from fitted parameters.** Polarized neutron reflectivity and polarization of Co/Ti on Ti substrate.

Figure 4 shows the PNR and polarization simulations using the fitted parameters of Co/Ti on Ti substrate in order to eliminate the direct beam and gaussian beam distribution to compare the theoretical simulations from Figure 1. Comparing the polarizations and reflectivities, Figure 4 confirms the theoretical simulations from Figure 1 and proves the concept of spin-down critical angle elimination and the maintenance of a relatively high spin-up critical angle by adjusting total SLD and the layer thickness ratio respectively. Such an improvement is appliable to not only this Bragg mirror geometry but also a supermirror geometry with a depth gradient layer thickness distribution. However, as mentioned, most importantly the possibility of extending the wavelength bandwidth according to Equation 1, to polarize very high wavelength neutrons in spallation sources.

We have confirmed the hypothesis and simulations with the experimental data, that the lower the average SLD of the nuclear SLD of the entire multilayer and substrate that the neutrons traverse through the lower the critical angle of spin down. We can expect the possibility to polarize at any angle above the critical angle for spin down as well, as long as the criteria for polarization is fulfilled, which it already is for the material system Co/Ti. By expanding the bandwidth of polarization in the lower scattering angles regions, the polarized neutron flux at neutron sources could be higher than state of the art, but most importantly, for a non-monochromatic beam, such as all the neutron spallation sources, meaning most of the neutron sources around the world, these optics would facilitate the polarization over a wavelength bandwidth several times higher than current state-of-the-art Fe/Si or Co/Ti on Si. Further the optics opens up the possibilities of polarizing neutrons with very large wavelengths.

# Methods

Co/Ti multilayer thin films were fabricated using ion-assisted DC magnetron sputter deposition in an ultra-high vacuum environment with a baseline pressure of approximately $2.6 \cdot 10^{-7}$ Pa ($2 \cdot 10^{-9}$ Torr). The deposition process used ion-assistance by attracting Ar-ions from the sputtering gas plasma using a substrate bias voltage of -80 V. A magnetic field, aligned with the substrate's normal, condensed the plasma closer to the substrate, enhancing the flux of Ar-ions arriving at the substrate. During sputtering, the substrate bias was first set to 0 V for the approximately initial 3 Å of each layer, whereafter a -80 V bias was applied for the remainder of the layer. This modulation in ion-assistance aimed to reduce layer intermixing.[19] Comprehensive details of the deposition system are presented in other sources.[20] The multilayers were deposited on Si(100) substrates, measuring $10 \times 10 \times 0.5$ mm$^3$, covered with native oxide. During deposition, the substrate was rotating at 15 rpm, with no deliberate heating applied. Targets of 50-mm-diameter were used for Co and Ti, and the deposition rates were determined by depositing multilayers with known deposition times for each layer and using X-ray reflectivity to determine the multilayer period thickness. The multilayered structures were made using computer-regulated shutters, controlling the flux of material from the magnetrons to achieve the intended layer thicknesses and designs. As for the buffer layer deposited before the multilayer for one of the samples, decreasing substrate bias voltages of -200 V, -150 V, and -100 V were applied sequentially for three 40 nm segments of the Ti buffer layer, totaling a 120 nm buffer layer thickness.

Hard X-ray reflectivity analysis was conducted using an Empyrean diffractometer from Panalytical. This was performed using a parallel beam setup with a line-focused copper anode source, operating at 45 kV and 40 mA, which produced Cu-K$\alpha$ radiation with a 1.54 Å wavelength. To condition the beam and ensure a specific X-ray spot size on the sample, a parabolic X-ray mirror and a ½° divergence slit were integrated into the incident beam path. For the diffracted beam, a parallel plate collimator paired with a 0.27° collimator slit was utilized, leading up to a PIXcel detector functioning in open detector mode. Multilayer periods were determined from the Bragg peak positions.

X-ray diffraction (XRD) studies were conducted using a Panalytical X'Pert diffractometer set up in Bragg-Brentano geometry. The system employed a Bragg-Brentano HD incident beam optics module, which incorporated a ½° divergence slit and a ½° anti-scatter slit. On the secondary optics side, a 5 mm anti-scatter slit was paired with an X'celerator detector operating in scanning line mode. The diffraction data collection spanned a 2θ range from 20° to 90°, with increments of 0.018° and a 20 s acquisition time per step.

Experiments using polarized neutron reflectivity (PNR) were conducted on the MORPHEUS reflectometer situated at SINQ, part of the Paul Scherrer Institut (PSI) in Villigen, Switzerland. In these experiments, an initially non-polarized neutron beam first interacts with a polarizer. This process ensures that only neutrons with a specific spin orientation continue. The beam's spin state, either up or down, can be toggled using a spin flipper. The polarized neutron beam is directed onto the sample surface, creating a specular reflection. The intensity of the reflection is influenced by how the beam interacts with all the sample's interfaces. PNR is effective in detecting the spin-dependent scattering length density (SLD) of the material, allowing for an in-depth study of its magnetic characteristics. Because of the two distinct spin orientations, PNR yields two separate reflectivity curves. The Bragg peaks that are observed are a result of constructive interference. During measurements, an external magnetic field, with an approximate strength of 20 mT, ensured that the samples were magnetically saturated in the in-plane direction. The reflective measurements spanned from 0 to 2° in 2θ and utilized a neutron wavelength of 4.83 Å.

## Acknowledgments

We are grateful to the Paul Scherrer Institute (PSI), Switzerland, for providing several beamtimes at MORPHEUS. Additionally, we would like to acknowledge the Swedish Research Council (VR) for their support under project numbers 2019-04837_VR (F.E.) and 2018-05190_VR (N.G). A.Z. acknowledge the support from the Hans Werthéns grant (2022-D-03), the Royal Academy of Sciences Physics grant (PH2022-0029), and the Lart Hiertas Minne grant (FO2022-0273).

# Appendix

## 1. Theory and equations

First in order to reflect neutrons, the scattering length density (SLD) of the two layers in a period needs to have a high contrast. The higher the contrast the higher the reflection. Secondly, neutrons can have two different spin states, namely spin up and spin down and in order to polarize the neutron so that only one spin state of neutrons is reflected while the other is not, the nuclear and magnetic SLD of the magnetic layer needs to be matched with the nuclear SLD of the non-magnetic layer. The most common material systems for polarizing neutron optics are Fe/Si and Co/Ti, both on Si substrates[2,10,11,13], seen in the sketches in Figure 1(a-b). Si substrates are the world's most common substrates for any thin film application and research due to its stability, low roughness, affluent single crystallinity, cost efficiency, etc.[21] Although previous research has extensively been focused to increase the polarization bandwidth to higher angles, meagre number of investigations have studied to polarize neutron below the critical angle.[8] Since the highest reflective intensity is below the critical angle, it would be very beneficial to polarize the non-polarized neutron beam below the critical angle.

The scattering length density (SLD) is given by the potential of a material:

$$V_0 = \frac{2\pi\hbar^2}{m}(SLD)$$

*(1)*

where V is the potential and ℏ is the reduced Planck's constant.[9] Further the neutron's kinetic energy kinetic energy that determines the reflection from the barrier can be written as:

$$E_{i\perp} = \frac{(k_i \hbar \sin\theta_i)^2}{2m}$$

*(2)*

where $k_i$ is the wave number and θ the incidence angle. Lastly, the scattering vector, which is proportionally dependent on the scattering angle, can be expressed as[22]:

$$q = \frac{4\pi}{\lambda}\sin\theta$$

*(3)*

where λ is the neutron wavelength. (1), (2) and (3) can be combined to obtain an expression for the critical angle:

$$q_c = \sqrt{16\pi(SLD)}$$

(4)

For samples with magnetic moments the critical angle may vary depending on the amplitude of the magnetic moment of a layer due to the magnetic moment of neutrons.[3] Since polarizing neutron optics are periodically magnetic multilayers, the spin up and spin down critical angles may differ depending on the magnetic moment, which in turn affects the magnetic SLD. For spin up neutrons the nuclear SLD (n-SLD) is added with the magnetic SLD (m-SLD) from the magnetic layer giving it a higher total SLD. While spin down neutrons subtract the m-SLD from the n-SLD, which leads to a lower total SLD. Since the $q_c$ is depending on the SLD not only from the top layer or layers but down to the substrate, also the substrate SLD matters to understand the value for the critical angle $q_c$. The thicker the film is the less does the substrate affects since due to the penetration depth of the neutrons.

## 2. X-ray diffraction

X-ray diffraction patterns seen in Figure 1 showcase the similarities of Co/Ti growth on the two different substrates Si and Ti. Where the Co stemming from the film is evident from the high and wide Co (111) peak, while the other peaks, only present for the Ti substrate sample, are stemming from the Ti (100), (002), (101) and (102) from the substrate itself and not the film. Using the FWHM of the Co (111), which is 0.78551° (/0.01371 rad) we calculate the crystallite size using the Scherrer equation to be 109Å which is well within the layer thickness being approximately 150Å.

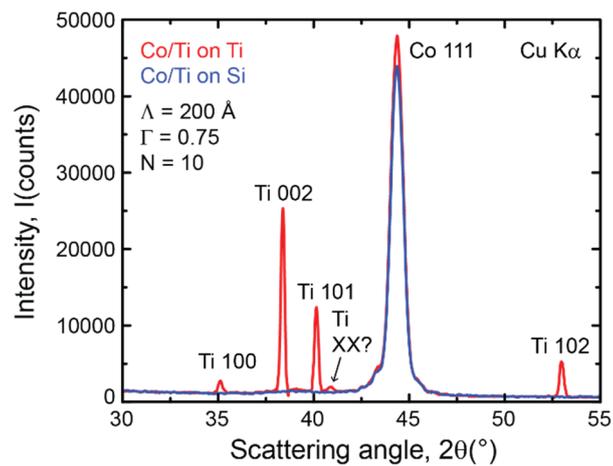

**Figure S1.** X-ray diffraction of Co/Ti multilayer on Si (brown) and Ti (blue). N = 10, Λ = 200 Å and Γ = 0.75.

# 3. X-ray reflectivity and reciprocal space mapping

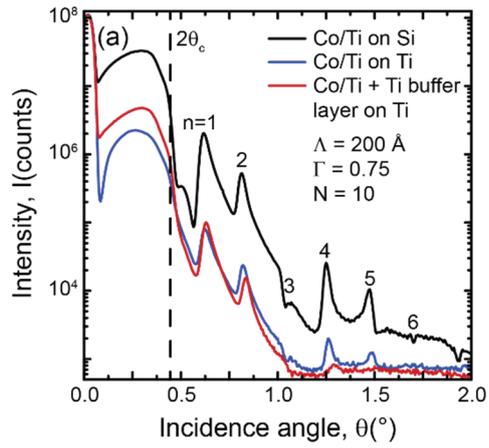

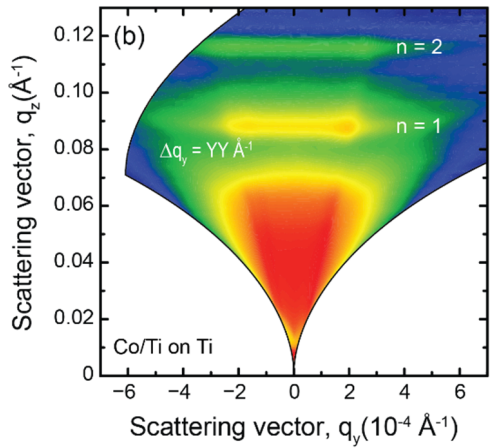

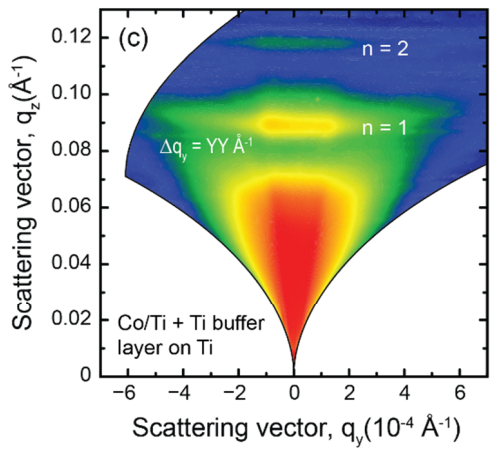

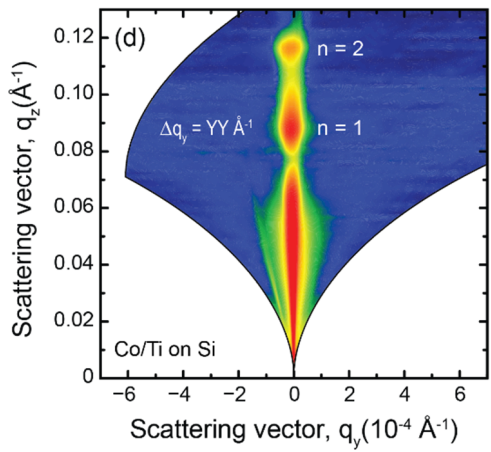

**Figure S2. X-ray reflectivity measurements.** (a) Comparison between Co/Ti on Si, Co/Ti on Ti substrate and Co/Ti + Ti buffer layer on Ti substrate (N = 10, $\Lambda$ = 200 Å and $\Gamma$ = 0.75 on all samples). (b-d) Off-specular X-ray reflectivity of the three samples from (a). The axes Qz and Qx represent the reciprocal space dimensions (Å$^{-1}$) common for off-specular scattering representation.

The multilayer growth on Ti substrate could cause difficulties since a Ti substrate as smooth as Si is an impossibility with current day techniques.[23] The Ti substrates used in this work were polished to the micrometer sized roughness level, and further polishing below this range was not feasible by the vendor, given Ti's inherent hardness. We propose and test a solution of depositing a thick buffer layer between the substrate and film,[24,25] before depositing the multilayer to combat current day rough Ti substrates. It is here of importance to use growth conditions that allow the Ti atoms to deposit more in the valleys of the roughnesses by for example applying high substrate bias voltages to increase the surface diffusion as done here, see Methods.

Figure S2(a) shows the X-ray reflectivity of three samples; Co/Ti on Si, on Ti and on a Ti buffer layer on top of a Ti substrate. The direct beam is shown from 0 to 0.2 degrees whereafter the region of total external reflection starts. The vertical grey lines represent the Bragg peak positions and shows that all three samples have the same period thicknesses. Figure S2(b-d) shows the off-specular X-ray scattering of the aforementioned three samples. The off-specular scattering maps use the Q-space (reciprocal space) representation for clarity.[26]

The X-ray reflectivity comparison in Figure S2(a) shows that the intensity from the critical edge down to the 4$^{th}$ order Bragg peak (6$^{th}$ if diminished Bragg peaks are taken into account) have an intensity difference between the sample Co/Ti on Ti (blue) and Co/Ti on Si (black) of an order of magnitude. The reflectivity notably diminishes for Co/Ti multilayers when deposited directly on Ti because of the micrometer-sized roughness on the Ti substrates. Further, the rounded shape of the critical edge for the Ti substrate sample is also evident of typical large micrometre sized roughness increasing the absorption which affects the total external reflection.[27] The multilayer deposited on 120 nm Ti buffer layer seem to have increased the critical angle and the 1$^{st}$ Bragg peak intensities as well as to shape the critical edge into a reflectivity curve more typical to smoother surfaces. The reason of the 2$^{nd}$, 3$^{rd}$ and 4$^{th}$ Bragg peak having a lower specular intensity is due to the increased accumulated nanometre scale roughness caused by the buffer layer. Hence more optimized growth conditions for smoother Ti layers could possibly increase even those Bragg peaks. Note also that 120 nm is not enough to compensate for the micrometre ranged roughnesses but experimentally leading towards the right direction by utilizing a Ti buffer layer. Demonstrated using off-specular X-ray reflectivity, seen in Figure S2(b-d) it is further evident how the off-specular scattering caused by the lateral roughness can be significantly decreased with the 120 nm Ti buffer layer, further proving the efficiency of our strategy. Using the width and height of the Bragg sheets the domain size ($\xi_d$) was calculated.[26,28] The values for the sample on Ti without buffer layer (b) and with buffer layer (c) were $\xi_d = 330$ nm and $\xi_d = 660$ nm, respectively. In other words, the domains are larger when the buffer layer was introduced, since smaller domains may have coalesced together, hence indicating that our strategy to get a net deposition mainly in the valleys seems to be efficient. Further, Ti is often used as an adhesion layer, so in this case the use of Ti as a buffer layer is not necessarily tricky in terms of buffer layer growth and adhesive strength between the substrate and film for industry either. As for the main comparison using neutrons between samples seen in Figure 1 (a, c and d), no buffer layer is needed for the comparison, since the comparison is for polarization and not as much reflectivity. The FWHM of the three samples at the first Bragg sheet are q = 0.2251, 0.1388 and 0.0194 nm$^{-1}$, respectively.

## 4. Polarized neutron reflectivity simulations

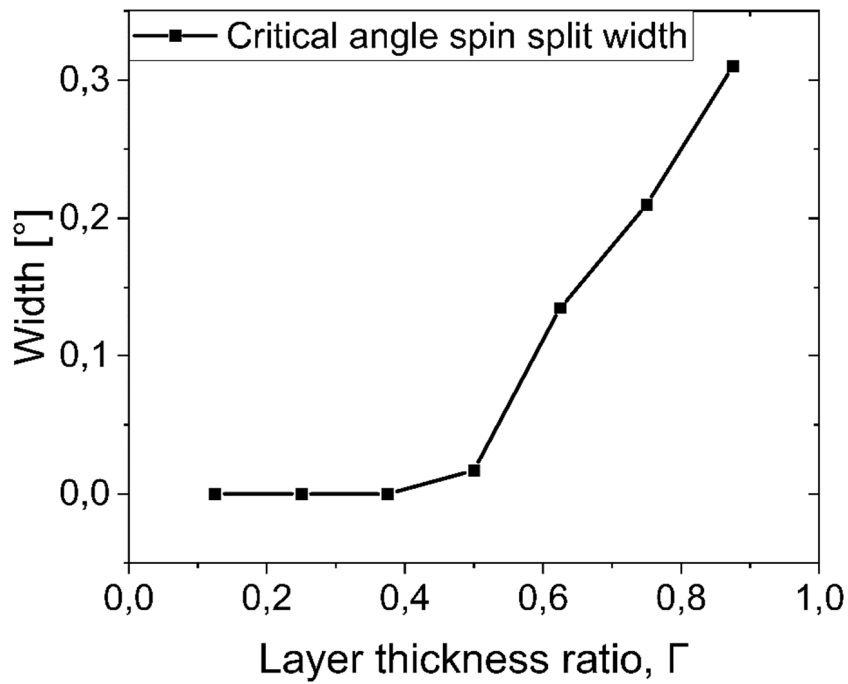

**Figure S3.** Critical angle spin split width simulations as a function of layer thickness ratio with parameters from simulation Figure 1.

Polarization is when there is a contrast between the two curves. In the case of neutron optics, one wants as high contrast as possible where the reflectivity is high as well. The higher the SLD the higher angle the critical edge will appear. Since spin-up neutrons are the nuclear SLD + magnetic SLD, while spin-down neutrons are nuclear SLD – magnetic SLD, the spin-up critical edge will be higher than the spin-down critical edge for Co/Ti or Fe/Si. Meaning that a high magnetic SLD will differentiate the angles of the critical angles the most. After the critical edge we see Kiessig fringes which occur due to the repetitions of the layers in the multilayer stack while the intense Bragg peak at around 2° occurs due to the constructive interferences that each period gave. Depending on the period thickness the Bragg peak would appear at different angles. In a traditional supermirror neutron optics it is this Bragg peak one wants to propagate by having an aperiodic multilayer with different thicknesses to achieve an extended region in angles with high reflectivity. For this study, a periodic multilayer is satisfactory, especially for investigating the critical angles where periodic or aperiodic does not matter.